\documentclass[prb,twocolumn,showpacs]{revtex4}

\usepackage{graphicx}

\newcommand{\ba}{\begin{eqnarray}}
\newcommand{\be}{\begin{equation}}
\newcommand{\ea}{\end{eqnarray}}
\newcommand{\ee}{\end{equation}}
\newcommand{\ds}{\displaystyle}

\newcommand{\SLS}{S_{\mathrm{LS}}}
\newcommand{\SHS}{S_{\mathrm{HS}}}
\newcommand{\sgn}{\mathrm{sgn}\,}
\newcommand{\Tr}{\mathrm{Tr}\,}

\newcommand{\ignore}[1]{}

\begin{document}

\title{Collective effects in spin-crossover chains with exchange interaction}

\author{Carsten Timm}
\email{ctimm@ku.edu}
\affiliation{Institut f\"ur Theoretische Physik, Freie Universit\"at Berlin,
Arnimallee 14, D-14195 Berlin, Germany\\
and Department of Physics and Astronomy, University of Kansas, Lawrence, Kansas
66045, USA}

\date{August 15, 2005}

\begin{abstract} The collective properties of spin crossover chains are
studied. Spin crossover compounds contain ions with a low-spin ground state
and low lying high-spin excited states and are of interest for molecular
memory applications. Some of them naturally form one-dimensional chains.
Elastic interaction and Ising exchange interaction are taken into account.
The transfer-matrix approach is used to calculate the partition function,
the fraction of ions in the high-spin state, the magnetization,
susceptibility etc.\ exactly. The high-spin/low-spin degree of freedom leads
to collective effects not present in simple spin chains. The ground state
phase diagram is mapped out and compared to the case with Heisenberg
exchange interaction. The various phases give rise to characteristic
behavior at nonzero temperatures, including sharp crossovers between low-
and high-temperature regimes. A Curie-Weiss law for the susceptibility is
derived and the paramagnetic Curie temperature is calculated. Possible
experiments to determine the exchange coupling are discussed. \end{abstract}


\maketitle

\section{Introduction}
\label{sec.intro}

Motivated in part by the search for molecular memory devices,
\cite{Kahn,BoK95,KaM98} spin-crossover compounds (SCCs) have been
investigated quite intensively in recent years. These compounds are
characterized by magnetic ions that can be either in a low-spin (LS) or
high-spin (HS) state, where the HS state is slightly higher in energy and
can be thermally populated.\cite{CaS31,Bonding1,Bonding2,GGG00,BlP04} An ion
in the HS (LS) state has the spin quantum number $\SHS$ ($\SLS$), where
$\SHS>\SLS$. Most SCCs consist of metal-organic complexes involving
transition-metal ions, in many cases $\mathrm{Fe}^{2+}$. 
Spin crossover involving charge transfer between
two different ions is observed in Prussian Blue
analogues.\cite{SIF96,GRN00,BLE00} SCCs are promising
for molecular memory applications due to long lifetimes of the HS state.
Applications require to deposit SCCs on substrates as thin films or
one-dimensional (1D) chains. Many SCCs consist of weakly interacting
chains or plains even in the bulk.\cite{KaM98,YMK98,Co,KhL04} A typical
quasi-1D material is $\mathrm{Fe}^{2+}$ with 4-R-1,2,4-triazole
ligands.\cite{KaM98,YMK98}

SCCs are also interesting from the point of view of basic physics. The
additional LS/HS degree of freedom leads to collective effects not present
in pure spin models, which are most pronounced for a diamagnetic LS state,
$\SLS=0$, as in the case of $\mathrm{Fe}^{2+}$ ions. Then, in the LS state
the spin is switched off. This reminds one of diluted spin
models,\cite{VMG02,dilute,YRH04} but with \emph{fluctuating} dilution.

Most of the theoretical literature on SCCs omits the exchange interaction.
The only interaction in this case is of \emph{elastic} origin and determines
whether neighboring ions prefer to be in the same (LS or HS) state or in
different ones.\cite{elastic,KhL04} However, typical values for the exchange
interaction $J$ in transition-metal complex salts are of the order of
$|J|/k_B = 10$ to $20\,\mathrm{K}$.\cite{WHM97,BSC04,LBP04} The interaction
is due to superexchange and is typically antiferromagnetic. In ferrimagnetic
Prussian Blue analogues the exchange interaction is typically
similar\cite{SIF96,GRN00} but values of the order of room temperature have
also been reported.\cite{FMO95}

If magnetic anisotropies are small, the exchange interaction is of
Heisenberg type. For strong easy-axis anisotropy we can instead consider
an Ising exchange interaction. For isolated 1D chains deposited on a surface
strong easy-axis anisotropy is expected due to \emph{shape anisotropy}.
Another important source of anisotropy is the ligand field acting on the
magnetic ions.
Nishino \textit{et al.}~\cite{NYM98,NiM01,NBM05} and
Boukhed\-da\-den \textit{et al.}~\cite{BNM05} study a three-dimensional
model with nearest-neighbor Ising exchange, $\SLS=0$, and $\SHS=1/2$ (the
high spins have two possible orientations) employing mean-field theory and
Monte Carlo simulations. On the other hand, Timm and
Schollw\"ock\cite{TiS05} consider a 1D model with nearest-neighbor
Heisenberg exchange, $\SLS=0$, and $\SHS=2$. The \emph{zero-temperature}
phase diagram is investigated using the density matrix renormalization
group.\cite{DMRG}

In this paper, we study the behavior of a 1D spin-crossover chain with and
without Ising exchange interaction at \emph{all} temperatures. We will argue
that its solution also provides a good approximation for the Heisenberg case for
small exchange interactions. In Sec.~\ref{sec:theory} we present the model and
discuss the transfer-matrix calculations. We mostly restrict ourselves to the
case $\SLS=0$ and $\SHS=2$, which is appropriate for $\mathrm{Fe}^{2+}$
compounds. The results are presented in Secs.~\ref{sec:ground} and
\ref{sec:finite} for the ground-state and finite-temperature properties,
respectively.

\section{Model and method}
\label{sec:theory}

This section outlines the theory
for general $\SHS$, unless stated
otherwise. We start from the Hamiltonian
\be
H = -V \sum_i \sigma_i \sigma_{i+1} - \Delta \sum_i \sigma_i
  - J \sum_i m_i m_{i+1} - h \sum_i m_i ,
\label{2.HI2}
\ee
where $m_i\equiv S_i^z$ with $m_i=0$ for $\sigma_i=1$ (LS state) and
$m_i=-\SHS,\ldots,\SHS$ for $\sigma_i=-1$ (HS state). Here, $V$ describes
the elastic interaction between neighboring spins.\cite{elastic} For $V>0$
($V<0$) homogeneous (alternating) arrangements of LS and HS are
favored.\cite{KhL04} $2\Delta$ is the energy difference between HS and LS
and $h$ is the magnetic field with a factor $g\mu_B$ absorbed. $J>0$ ($J<0$)
is a ferromagnetic (antiferromagnetic) exchange coupling. For $J=0$ this
model is equivalent to the one of Ref.~\onlinecite{GKA00} and
for $J=h=0$ it is the model introduced by Wajn\-flasz and Pick\cite{WaP70}
and by Doniach.\cite{Don78}

We apply the transfer-matrix approach. To focus on the essential physics we
assume that the vibrational frequencies of a complex do not change between
the LS and HS state. Then the degeneracy of the HS state is only due to its
spin. The main effects of different vibrational frequencies would be to
renormalize the energy splitting $2\Delta$ and to change the effective
degeneracies of the LS and HS states.\cite{BoK95,GKA00} These effects do not
change the qualitative results and can easily be reintroduced.

The total partition function is
$Z = \sum_{\{\sigma_i,m_i\}} e^{-\beta H}$,
where $\beta=1/T$ (we set $k_B=1$). We write
$-\beta H \equiv \sum_i K_{\sigma_im_i,\sigma_{i+1}m_{i+1}}$
with
\be
K_{\sigma m,\sigma' m'} = \beta V \sigma \sigma'
  + \frac{\beta\Delta}{2}\,
    (\sigma+\sigma')
  + \beta J m m'
  + \frac{\beta h}{2}\, (m + m') .
\ee
The partition function then reads
\be
Z = \sum_{\{\sigma_i,m_i\}} e^{K_{\sigma_1 m_1,\sigma_2 m_2}}
  e^{K_{\sigma_2 m_2,\sigma_3 m_3}} \cdots
  e^{K_{\sigma_N m_N,\sigma_1 m_1}} .
\ee
This can be written as
$Z = \Tr M^N$ where the symmetric matrix $M$ has the components
$M_{\sigma m,\sigma' m'} = \exp(K_{\sigma m,\sigma' m'})$.
For $N\to\infty$ the partition function
becomes the maximum eigenvalue of $M$ to the power $N$ and the 
Gibbs' free energy per site is
\be
g = -T \lim_{N\to\infty} \frac{1}{N} \ln \Tr M^N .
\ee
For $J=0$ the
eigenvalue equation can be solved in closed form. For $\SLS=0$ we obtain
\ba
g & = & -\frac{1}{\beta} \ln \left( e^{\beta V} \cosh\beta \tilde\Delta
  + \sqrt{e^{2\beta V} \sinh^2\beta \tilde\Delta + e^{-2\beta V}}\right)
  \nonumber \\
& & {}-\frac{\ln g_{\mathrm{HS}}}{2\beta}
\label{2.g3}
\ea
with $\tilde\Delta \equiv \Delta - {\ln g_{\mathrm{HS}}}/{2\beta}$
and\cite{GKA00}
\be
g_{\mathrm{HS}} \equiv \sum_{m=-\SHS}^{\SHS} e^{\beta h m}
  = \frac{e^{\beta h (\SHS+1/2)}-e^{-\beta h (\SHS+1/2)}}
  {e^{\beta h/2}-e^{-\beta h/2}} .
\label{2.gdef1}
\ee
For $h=0$ one has $g_{\mathrm{HS}}=2\SHS+1$.

For $J\neq 0$, $\SLS=0$, $\SHS=2$ the calculation of the partition function
has been reduced to the eigenvalue problem of the $6\times 6$ matrix $M$. An
important quantity describing SCCs is the fraction $\gamma$ of ions in the
HS state. It is given by $\gamma = (1-\langle\sigma\rangle)/2$ so that, with
Eq.~(\ref{2.g3}),
\be
\gamma = \frac{1+\partial g/\partial \Delta}{2}
  = \frac{1}{2} - \frac{T}{2} \lim_{N\to\infty}
  \frac{\Tr M^{N-1}\, \partial M/\partial \Delta}{\Tr M^N} .
\ee
If $|n\rangle$, $n=1,\ldots,6$ are the orthonormalized eigenvectors of
$M$ with eigenvalues $m_1\ge m_2\ge \cdots \ge m_6$, then
\be
\gamma = \frac{1}{2} - \frac{T}{2}\, \frac{\langle 1| \partial M/
  \partial \Delta|1\rangle}{m_1} .
\ee
Similarly, the magnetization is
\be
\langle m\rangle = -\frac{\partial g}{\partial h}
  = T\, \frac{\langle 1| \partial M/
  \partial h|1\rangle}{m_1}
\label{2.avmJ3}
\ee
and the probability of any two neighbors
being in the same (LS or HS) state is
\be
w_{\mathrm{eq}} = \frac{1-\partial g/\partial V}{2}
  = \frac{1}{2} + \frac{T}{2}\, \frac{\langle 1| \partial M/
  \partial V|1\rangle}{m_1} .
\ee
This quantity describes nearest-neighbor correlations.

For the special case $J=0$ we find the HS fraction
\be
\gamma = \frac{1}{2} - \frac{1}{2}\, \frac{e^{\beta V} \sinh\beta
  \tilde\Delta}{\sqrt{e^{2\beta V}\sinh^2\beta \tilde\Delta
  + e^{-2\beta V}}} .
\label{2.gamma3}
\ee
In this case the magnetization is determined by $\gamma$
through $\langle
m\rangle = \gamma \SHS B_{\SHS}(\beta h\SHS)$, where
$B_S(x)$ is the
Brillouin function. This leads to a simple result for
the susceptibility $\chi = \partial\langle m\rangle/\partial h$
for vanishing magnetic field:
\be
\chi = \gamma\,\frac{\SHS(\SHS+1)}{3T} .
\label{2.chiex2}
\ee
Thus the susceptibility is just the paramagnetic expression weighted by the
concentration of high spins.

\begin{figure}[t]
\centerline{\includegraphics[width=3.20in]{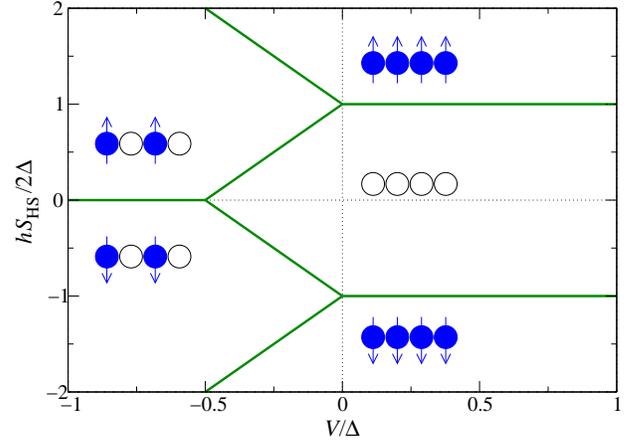}}
\caption{(Color online) Ground-state phase diagram of the
spin-crossover chain without exchange interaction and with $\SLS=0$, in
terms of elastic interaction $V$ and Zeeman energy $h$ in units of $\Delta$,
where $2\Delta$ is the LS/HS energy splitting. The solid lines denote
discontinuous transitions.}
\label{fig.PD0}
\end{figure}

\begin{figure}[tbh]
\centerline{\includegraphics[width=3.20in]{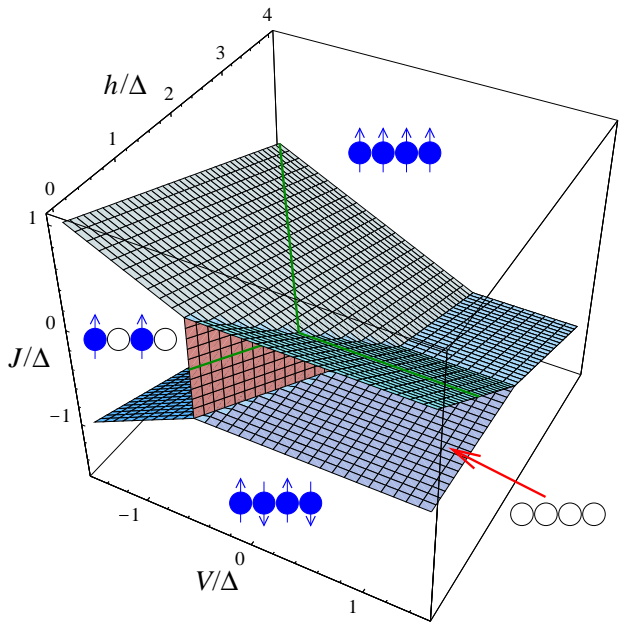}}
\caption{(Color online)
Ground-state phase diagram of the spin-crossover chain
with Ising exchange coupling $J$. In this and the following graphs
$\SLS=0$ and $\SHS=2$ are assumed.
The planes denote discontinuous transitions.
The heavy solid lines show their intersections
with the $J=0$ plane, cf.\ Fig.~\protect\ref{fig.PD0}.}
\label{fig.PD4}
\end{figure}

\section{Ground states}
\label{sec:ground}

We now derive the ground-state phase diagram to lay the ground
for the discussion of finite-temperature properties. For the case without
exchange interaction Fig.~\ref{fig.PD0} shows the phase diagram in
terms of the two dimensionless ratios $V/\Delta$ and $h/\Delta$. From
Eq.~(\ref{2.g3}) we find $\gamma=0$, $\langle m\rangle=0$, and
$w_{\mathrm{eq}}=1$ for $|h|/\Delta<2/\SHS$ and $V/\Delta>-1/2 +
|h|\SHS/4\Delta$. This is the LS state. For $|h|/\Delta>2/\SHS$ and
$V/\Delta>1/2 - |h|\SHS/4\Delta$ we find $\gamma=1$, $\langle
m\rangle=\SHS\,\sgn h$, and $w_{\mathrm{eq}}=1$. This is the fully polarized
HS state. In all other cases we find $\gamma=1/2$, $\langle m\rangle =
\SHS\, \sgn h/2$, and $w_{\mathrm{eq}}=0$, corresponding to an
\emph{alternating} state of LS and HS ions.

\begin{figure}[tbh]
\centerline{\includegraphics[width=3.20in]{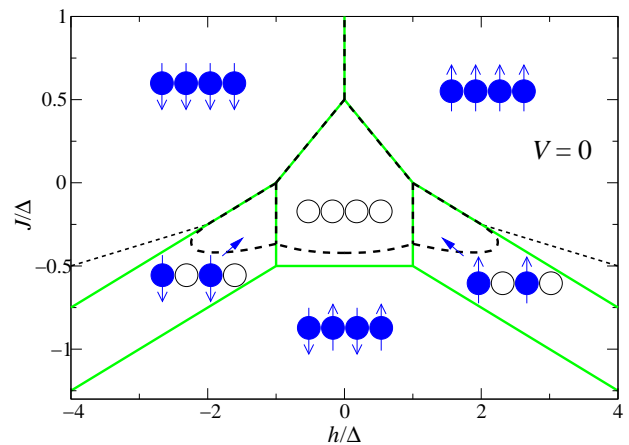}}
\caption{(Color online) Solid lines: Intersection of the
$T=0$ phase diagram, Fig.~\protect\ref{fig.PD4}, with the
$V=0$ plane (vanishing elastic interaction). Dashed curves:
The phase boundaries for \emph{Heisenberg} exchange
interaction.\protect\cite{TiS05} Long-dashed (short-dashed) curves correspond
to discontinuous (continuous) transitions. The alternating phase is reduced
to two finite lobes by quantum effects.
Note that the phase boundaries for the Ising- and Heisenberg-type models
are identical for positive and small negative $J$.}
\label{fig.PD5}
\end{figure}

In the general case with Ising exchange interaction we have three
dimensionless parameters, $V/\Delta$, $J/\Delta$, and $h/\Delta$. The ground
state is found by determining the largest component(s)
of the matrix $M$, since for $T\to 0$ these components become exponentially
larger than the others. For sufficiently
large $J/\Delta>0$ we obtain a ferromagnetically aligned HS phase for
any $V$, $h$. For $\SHS=2$ and $h\ge 0$ this phase becomes unstable at
\be
\hspace*{-0.5em}
\frac{J}{\Delta} = \left\{\!\begin{array}{ll}
\ds -\frac{h}{2\Delta} + \frac{1}{2} & \ds\mbox{for } \frac{h}{\Delta}<2
  \mbox{ and } \frac{V}{\Delta} > \frac{h}{2\Delta} - \frac{1}{2} \\[1.5ex]
\ds -\frac{V}{2\Delta} - \frac{h}{4\Delta} + \frac{1}{4} & \!\!
  \left\{\!\!\begin{array}{l}
  \ds\mbox{for }
    \frac{h}{\Delta}<2 \mbox{ and } \frac{V}{\Delta} < \frac{h}{2\Delta}
    - \frac{1}{2}
    \\[1.5ex]
  \ds\mbox{for } \frac{h}{\Delta}>2 \mbox{ and } \frac{V}{\Delta}
    < \frac{1}{2_{}}
  \end{array}\right. \\[1.5ex]
\ds -\frac{h}{4\Delta} & \ds\mbox{for } \frac{h}{\Delta}>2 \mbox{ and }
  \frac{V}{\Delta} >
  \frac{1}{2} .
\end{array}\right. \hspace*{-1em}
\ee
For large $-J/\Delta>0$ we find a HS phase with \emph{N\'eel order}, which
does not exist for $J=0$. It becomes unstable at
\be
\frac{J}{\Delta} = \left\{\!\begin{array}{ll}
\ds -\frac{1}{2} & \ds\mbox{for } \frac{h}{\Delta} < 2 \mbox{ and }
  \frac{V}{\Delta} > \frac{h}{2\Delta} - \frac{1}{2} \\[1.5ex]
\ds \frac{V}{2\Delta} - \frac{h}{4\Delta} - \frac{1}{4} & \!\!
  \left\{\!\!\begin{array}{l}
  \ds\mbox{for }
    \frac{h}{\Delta}<2 \mbox{ and } \frac{V}{\Delta} < \frac{h}{2\Delta}
    - \frac{1}{2}
    \\[1.5ex]
  \ds\mbox{for } \frac{h}{\Delta}>2 \mbox{ and } \frac{V}{\Delta}
    < \frac{1}{2_{}}
  \end{array}\right. \\[1.5ex]
\ds -\frac{h}{4\Delta} & \ds\mbox{for } \frac{h}{\Delta}>2 \mbox{ and }
  \frac{V}{\Delta} > \frac{1}{2} .
\end{array}\right.
\ee
Finally, for small $|J/\Delta|$ the LS and alternating phases are
separated by a boundary at
\be
\frac{V}{\Delta} = \frac{h}{2\Delta} - \frac{1}{2} \quad
  \mbox{for } -\frac{1}{2} < \frac{J}{\Delta} < \frac{1}{2} -
  \frac{h}{2\Delta} .
\ee
The resulting $T=0$ phase diagram is shown in Fig.~\ref{fig.PD4}.
Figure~\ref{fig.PD5} shows the $V=0$ cut of the
phase diagram and compares the results to the case of \emph{Heisenberg}
exchange interaction from Ref.~\onlinecite{TiS05}.
In the Heisenberg-type model several more complex phases are found, which
are absent in the Ising case. These consist of arrangements of 2, 3, 5, 7,
or 9 consecutive ions in the HS state followed by a single ion in the LS
state. This pattern is periodically repeated.\cite{TiS05} These phases are
stabilized in the Heisenberg case by the energy gain for
antiferromagnetically coupled spins due to quantum effects. This energy gain
also stabilizes the antiferromagnetically aligned HS state, which displaces
the other phases, as seen in Fig.~\ref{fig.PD5}.
%

On the other hand, for all positive and small negative $J/\Delta$ the phase
boundaries coincide. This is because the LS, alternating, and ferromagnetic
HS ground states of the Ising-type model remain eigenstates, with the same
energy, of the Heisenberg-type model. (The LS and alternating states
do not contain nearest-neighbor HS pairs so that the exchange interaction
does not enter. The ferromagnetically fully aligned state is an eigenstate
for any---even antiferromagnetic---Heisenberg-type model.)
In addition, all ground states that appear at larger $|J|$ have nonzero
energy gaps to the ground state at small $|J|$. Thus for small $|J|$, which
are expected for most SCCs, no new phases appear
for either the Ising or the Heisenberg case and both have the same ground
states.

\section{Finite-temperature behavior}
\label{sec:finite}

We now turn to the finite-temperature properties.
The partition function is analytic for $T>0$ so that the
phase transitions are replaced by cross\-overs. These can
become very sharp, however.

While the assumption of Ising exchange is appropriate in the
presence of strong easy-axis anisotropy, it also provides a
reasonable approximation for the Heisenberg case for small $J$. In this
case the ground state is found exactly and the
energies of low-lying excited states above the LS and alternating ground
states are also identical for both cases so that the behavior at low
temperatures will be very similar. Deviations will appear when a significant
fraction of nearest neighbors are both in the HS state since then the
exchange interaction becomes relevant. We will come back to
this at the end of this section. On the other hand, in the ferromagnetic HS
state the low-energy excitations are different but gapped in both cases.
Thus only qualitatively similar behavior is expected.
%

Note that for solving the Heisenberg-type model,
mapping onto and solving an Ising-type model is superior
to the mean-field approximation, which neglects all fluctuations.
In the Ising approximation we treat
fluctuations of $\sigma_i$ and \emph{longitudinal} fluctuations of
$\mathbf{S}_i$ exactly and neglect only \emph{transverse} fluctuations of
$\mathbf{S}_i$.

\subsection{Low-spin phase}

\begin{figure}[t]
\centerline{\includegraphics[width=3.20in,clip]{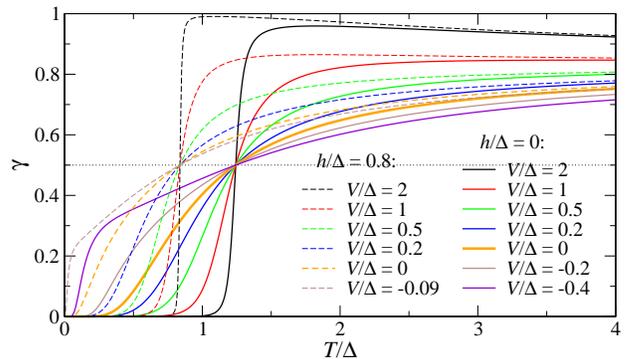}}
\caption{(Color online)
Solid curves: HS fraction $\gamma$ as a function of
temperature for various values of the elastic interaction $V$ and
$J=h=0$. Dashed curves: $\gamma$ for $h/\Delta=0.8$ and $J=0$.
The ground state is in the LS phase.}
\label{fig.gammaLS}
\end{figure}

Typical SCCs are in the LS phase for $T\to 0$. Figure \ref{fig.gammaLS}
shows the HS fraction as a function of temperature for $J=0$ and various
elastic interaction strengths $V$. For $V=0$ we observe the well-known
smooth crossover to the HS state, which comes from its higher degeneracy.
For $T\to\infty$ the HS fraction is only determined by the degeneracies,
$\gamma\to (2\SHS+1)/(2\SHS+2) = 5/6$, for any $V$. For $V>0$
the interaction favors a homogeneous state and the crossover becomes
\emph{sharper}. In \emph{mean-field theory} we would eventually reach a
critical point $V=V_c$ and find a discontinuous transition for
$V>V_c$.\cite{KaM98}
However, the exact solution for 1D shows that
$\gamma$ develops a discontinuity only for
$V\to\infty$.
For larger $V/\Delta$ the curves \emph{overshoot} and exhibit
a maximum. This is due to the energy gain from the homogeneous HS state
outweighing the entropy gain from populating all states equally. For $V<0$
the interaction favors a mixed state
with $\gamma\approx 1/2$ and thus broadens the crossover.

\begin{figure}[t]
\centerline{\includegraphics[width=3.20in,clip]{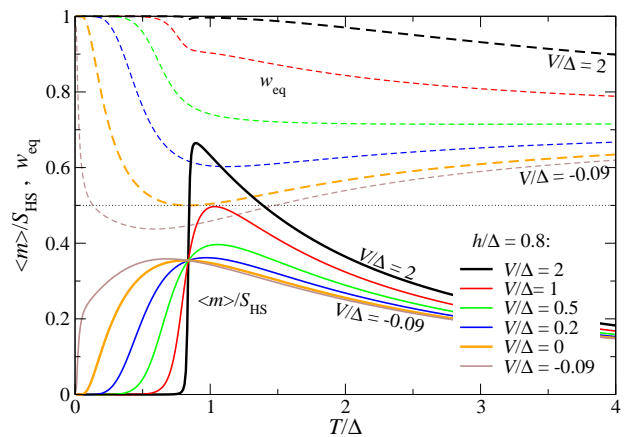}}
\caption{(Color online)
Solid curves: Magnetization $\langle m\rangle/\SHS$ as a function of
temperature for various values of the elastic interaction $V$, Zeeman energy
$h/\Delta=0.8$, and exchange $J=0$. Dashed curves of same color: Probability
$w_{\mathrm{eq}}$ of neighboring ions being in the same (LS or HS) state for
the same parameters. The ground state is in the LS phase.}
\label{fig.magLS}
\end{figure}

All curves for fixed $h$ in Fig.~\ref{fig.gammaLS} cross at $\gamma=1/2$.
The equation $\gamma(T_\gamma)=1/2$ is equivalent to
$\langle\sigma\rangle=0$ and, using Eq.~(\ref{2.gamma3}), to
$\Delta-({T}/{2})\,\ln g_{\mathrm{HS}}=0$. This leads to the implicit
equation ${T_\gamma}/{\Delta} = {2}/{\ln g_{\mathrm{HS}}(h,T_\gamma)}$,
which is indeed independent of the interaction $V$ so that all curves cross
at $T_\gamma$. For $h=0$ we obtain $T_\gamma/\Delta = 2/\ln (2\SHS+1)$.

The magnetization $\langle m\rangle$ and probability $w_{\mathrm{eq}}$ for
equal neighbors are shown in Fig.~\ref{fig.magLS} for $h/\Delta=0.8$. For
$V=0$ we see a broad crossover in $\langle m\rangle$ from the LS phase to a
maximum and then a decay $\sim 1/T$ at high temperatures. The Curie-Weiss
law is discussed further in Sec.~\ref{sec:high}. For $V>0$ ($V<0$) the
crossover becomes sharper (broader). Similar behavior is found for $J\neq 0$.
The nonmonotonic temperature dependence can be interpreted as the 1D
analogue of the reentrant transition to ferromagnetic order found in three
dimensions.\cite{BNM05}

The probability $w_{\mathrm{eq}}$ starts from $w_{\mathrm{eq}}=1$ at $T=0$
and approaches a universal value of $13/18$ for $T\to\infty$.
$w_{\mathrm{eq}}$ shows a broad minimum for small $V$, which is best
understood in the case without interactions. Then $w_{\mathrm{eq}} =
\gamma^2 + (1-\gamma)^2$ so that $\gamma=1/2$ implies $w_{\mathrm{eq}}=1/2$,
which is smaller than the limits for low and high $T$.

\begin{figure}[t]
\centerline{\includegraphics[width=3.20in,clip]{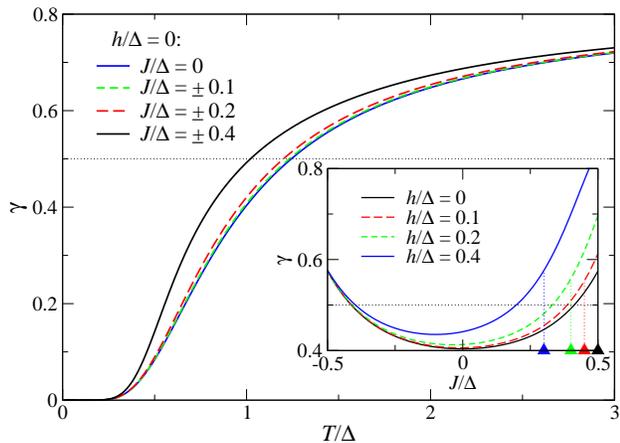}}
\caption{(Color online) HS fraction $\gamma$ as a function of
temperature for various values of the exchange interaction $J$ and
$V=h=0$. The ground states are always in the LS phase.
Inset: HS fraction $\gamma$ as a function of
exchange interaction $J/\Delta$ for various values of the Zeeman energy
$h$ for $T/\Delta=1$ and $V=0$.
The $T=0$ transition between LS and antiferromagnetic HS phases it at
$J/\Delta=-1/2$, the positions of the transitions to the
ferromagnetic HS phase are marked by triangles of the same color as the
corresponding curves.}
\label{fig.JgammaLS}
\end{figure}

Next, we consider the effect of the exchange interaction, see
Fig.~\ref{fig.JgammaLS}. For vanishing magnetic field the effect of moderate
$|J/\Delta|$ is rather weak due to the scarcity of nearest-neighbor HS
pairs. For $J\neq 0$ the crossover temperature with $\gamma(T_\gamma)=1/2$
depends on both $J$ and $V$. $\gamma$ increases \emph{symmetrically} for
fer\-ro\-mag\-ne\-tic and antiferromagnetic $J$ for all $V$, due to the
invariance of the Hamiltonian under $m_i\to (-1)^i m_i$ and $J\to-J$ for
$h=0$. The $T=0$ transitions to the ferromagnetic and antiferromagnetic HS
phases take place at $J/\Delta=\pm 1/2$, respectively. For $h\neq 0$ the
transition to the ferromagnetic phase shifts to lower values of $J$, whereas
the antiferromagnetic transition remains fixed, see Fig.~\ref{fig.PD5}, so
that the minimum of $\gamma(J)$ shifts to lower $J$, as seen in the inset of
Fig.~\ref{fig.JgammaLS}.

\begin{figure}[tbh]
\centerline{\includegraphics[width=3.20in,clip]{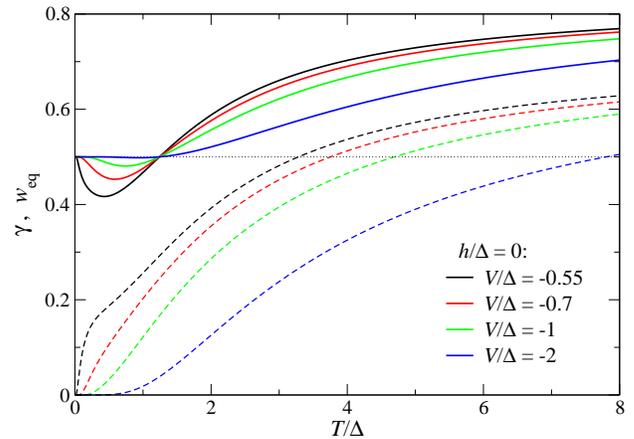}}
\caption{(Color online) HS fraction
$\gamma$ (solid curves) and probability $w_{\mathrm{eq}}$ of neighbors
in the same state (dashed curves of same color) as functions of
temperature for various values of the elastic interaction $V$
and $J=h=0$. The ground state is in the alternating phase.}
\label{fig.gammaLSHS}
\end{figure}

\subsection{Other phases}

The alternating LS/HS phase is stabilized by a strongly negative elastic
interaction $V$.\cite{KhL04} This can be expected for SCCs grown on a
substrate, where the transition of an ion to the HS state increases its
radius, which should favor a smaller ionic radius (LS state) of its
neighbors.

Figure \ref{fig.gammaLSHS} shows the HS fraction and the probability of
neighbors in the same state for $J=h=0$. $\gamma$ first decreases with
temperature and shows a universal crossing at $\gamma=1/2$ and $T=T_\gamma$
as in the LS phase [$T_\gamma/\Delta=2/\ln(2\SHS+1)$ for $h=0$].
The HS fraction first decreases since the alternating
order is partially destroyed ($w_{\mathrm{eq}}$ increases) and the energy of
spins having one LS and one HS neighbor is only determined by $\Delta$,
which favors the LS state. Since the effect of $J\neq 0$
is weak as long as the ground state is not changed,
we do not show the $J$ dependence here. At zero field
$\gamma$ is an even function of $J$, as for the LS phase.

\begin{figure}[t]
\centerline{\includegraphics[width=3.20in,clip]{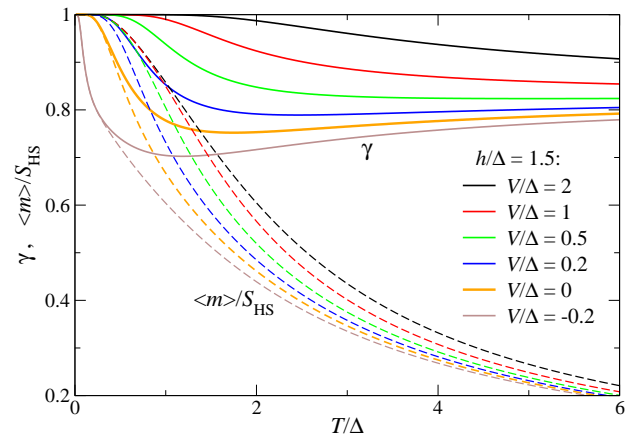}}
\caption{(Color online) HS fraction $\gamma$ (solid curves)
and magnetization $\langle m\rangle$ (dashed curves of same color)
as functions of
temperature for various values of the elastic interaction $V$,
$J=0$, and $h/\Delta=1.5$. The state at
$T=0$ is in the HS phase.}
\label{fig.gammaHS}
\end{figure}

While the exchange interaction has only a weak effect in the LS and
alternating phases, the situation changes for the HS phases. The
\emph{ferromagnetically} aligned phase appears at (very strong) magnetic
fields and for strong ferromagnetic exchange interaction---both might not be
realized in SCCs. Typical results for the \emph{field-induced} ferromagnetic
HS state are shown in Fig.~\ref{fig.gammaHS}. $\gamma$ goes through a
minimum since the ground state of a single ion would be the lowest state of
the HS quintet, while the first exited state is the LS singlet. Therefore
for increasing temperature the LS state contributes first. The magnetization
for $J=0$ is determined by $\gamma$ and a Brillouin function containing the
external field, as noted in Sec.~\ref{sec:theory}.

\begin{figure}[t]
\centerline{\includegraphics[width=3.20in,clip]{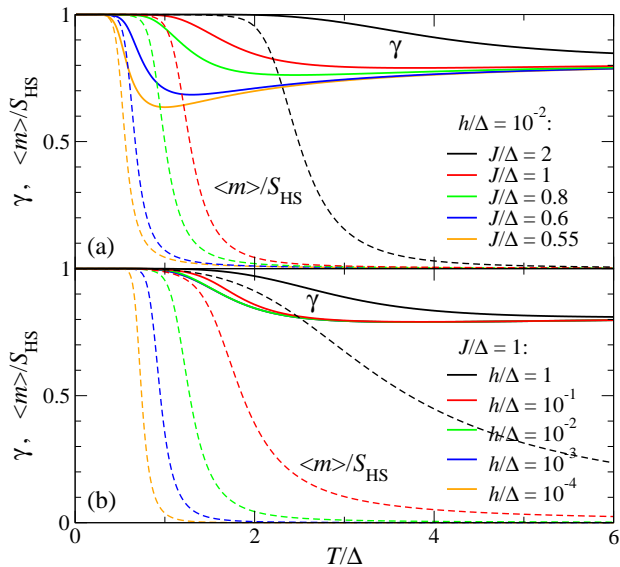}}
\caption{(Color online) (a) HS fraction $\gamma$ (solid curves) and
magnetization $\langle m\rangle$ (dashed curves of same color) as functions
of temperature for various values of the exchange interaction $J$, $V=0$,
and small $h/\Delta=10^{-2}$. (b) The same for fixed
$J/\Delta=1$ and various Zeeman energies. The state at $T=0$ is in the
ferromagnetic HS phase.}
\label{fig.JgammaHS}
\end{figure}

For the ferromagnetic phase induced by \emph{strong exchange} at small
field, Fig.~\ref{fig.JgammaHS} shows typical results for the HS fraction and
magnetization. $\gamma$ and $\langle m\rangle$ now depend \emph{strongly} on
$J$. The HS fraction behaves similarly to the case of $J=0$ and large field
but now the simple relation between $\gamma$ and $\langle m\rangle$ no
longer holds---it would predict a tiny magnetization at $h/\Delta=10^{-2}$.
Instead, the magnetization shows a rather sharp crossover from nearly full
polarization to low magnetization. Figure \ref{fig.JgammaHS}(b) shows the
dependance of $\gamma$ and $\langle m\rangle$ on magnetic field. At small
$h/\Delta\lesssim 10^{-2}$ the HS fraction hardly depends on $h$, but the
magnetization shows a strong field dependence. There is a strong tendency
towards ferromagnetic order even at very small $h$, whereas the
magnetization vanishes for $h=0$. We return to this crossover in
Sec.~\ref{sec:cross}.

\begin{figure}[tbh]
\centerline{\includegraphics[width=3.20in,clip]{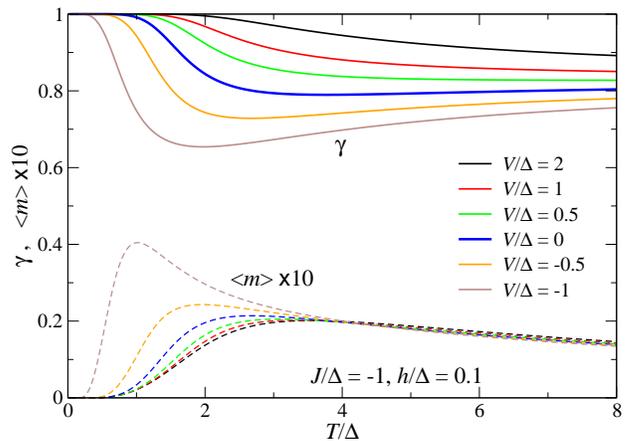}}
\caption{(Color online) High-spin fraction $\gamma$ (solid curves) and
magnetization $\langle m\rangle$ (dashed curves of same color, scaled by
$10$) as functions of temperature for various values of the elastic
interaction $V$, $h/\Delta=0.1$, and $J/\Delta=-1$.
The state at $T=0$ is in the antiferromagnetic HS phase.}
\label{fig.JgammaAFM}
\end{figure}

Finally, the \emph{antiferromagnetically} aligned HS phase only exists for
strong antiferromagnetic exchange $J/\Delta<-1/2$. Figure
\ref{fig.JgammaAFM} shows the HS fraction and the magnetization for fixed
$J/\Delta=-1$ in a magnetic field. $\gamma$ behaves very similarly to the
ferromagnetic phase, Fig.~\ref{fig.gammaHS}. The field induces a nonzero
magnetization only for $T>0$ because of the energy gap for spin flips. For
Heisenberg exchange the temperature dependence would be quite different: The
ground state in that case still has $\gamma=1$ but no long-range order due
to transverse spin fluctuations. However, it is also gapped since $\SHS$ is
an integer.\cite{Hal83}

\subsection{High-temperature limit}
\label{sec:high}

In the present subsection we consider the high-tem\-pe\-ra\-ture expansion
of thermodynamic quantities. We discuss the Curie-Weiss law for the
susceptibility and possible experiments to determine the exchange coupling.

\begin{figure}[t]
\centerline{\includegraphics[width=3.20in,clip]{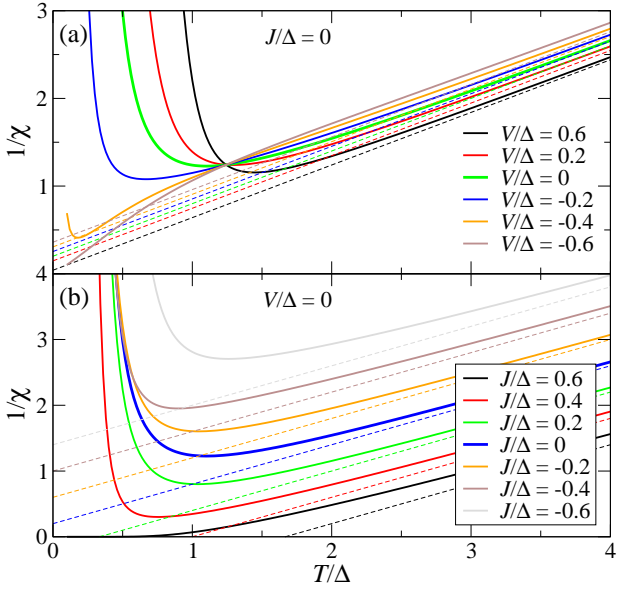}}
\caption{(Color online) Inverse susceptibility $1/\chi$
as a function of temperature for vanishing magnetic field, $h=0$.
(a) For $J=0$ and various values of $V/\Delta$. (b)
For $V=0$ and various values of $J/\Delta$. The solid curves show the exact
results, while the dashed curves of the same color give the Curie-Weiss law
$\chi \protect\cong 5/(3(T-\Theta))$ with $\Theta$ given by
Eq.~(\protect\ref{3.Theta2}).}
\label{fig.chi}
\end{figure}

In Sec.~\ref{sec:theory} the partition function $z$ for a single
site has been obtained as the largest eigenvalue of the matrix $M$.
We write $M = M_\infty + \Delta M$ with
$(M_\infty)_{\sigma m,\sigma'm'} \equiv \lim_{T\to\infty}
  M_{\sigma m,\sigma'm'} = 1$
and calculate $z$
perturbatively for small $\Delta M$. $M_\infty$ has an eigenstate
$|1\rangle_0=(1,1,1,1,1,1)/\sqrt{6}$ with eigenvalue $2\SHS+2=6$ and all other
eigenvalues vanish. Using standard perturbation theory for $z$ and the
corresponding eigenstate $|1\rangle$ and expanding in $1/T$ we obtain
\ba
\lefteqn{z \cong 6 + \frac{8V - 12\Delta}{3T} } \nonumber \\
& & {}+ \frac{121 V^2 - 168 V\Delta + 81 \Delta^2 + 225 J^2 + 135 h^2}
  {27 T^2}
  \nonumber \\*
& & {}+ \big(148 V^3 - 846 V^2\Delta + 774 V\Delta^2 + 2925 VJ^2 \nonumber \\
& & \quad{}+ 1080 Vh^2 - 162 \Delta^3 - 2700 \Delta J^2 - 1215 \Delta h^2
  \nonumber \\
& & \quad{}+ 4050 Jh^2\big)/\big(243 T^3\big) + {\cal O}(1/T^4) .
\label{3.z2}
\ea
The Gibbs' free energy per site is then $g=-T\ln z$
and the susceptibility is
\be
\chi = -\frac{\partial^2 g}{\partial h^2}
  \cong \frac{5}{3\,T} + \frac{5(4V - 3\Delta + 30J)}{27\, T^2} + {\cal O}(1/T^3) .
\ee
Expanding the Curie-Weiss law $\chi = C/(T-\Theta)$ we find
\be
\chi \cong \frac{5}{3}\, \frac{1}{T-\Theta}
\ee
with the paramagnetic Curie temperature
\be
\Theta = \frac{4V-3\Delta+30J}{9} .
\label{3.Theta2}
\ee
Figure \ref{fig.chi} shows the approach of the exact result for $1/\chi$
to the Curie-Weiss form.

We see that the high-temperature behavior of the susceptibility not only
depends on the exchange interaction $J$ but also  on the elastic interaction
and the LS/HS energy splitting. Interestingly, the paramagnetic Curie
temperature $\Theta$ is nonzero even without \emph{any} interaction, in
which case $\Theta = -\Delta/3$. Thus there is a deviation from the Curie
law for the susceptibility of noninteracting spin-crossover ions. The
reason is that the ground state is diamagnetic and thermal activation with
$T\sim \Delta$ is required to make an ion paramagnetic. Obviously,
$\Theta<0$ does not imply an antiferromagnetic ground state. Similarly, the
ground state is not always ferromagnetic if $\Theta>0$---the tendency
towards ferromagnetism can be preempted by the crossover to the LS state.
This is seen for $J/\Delta=0.2$ and $0.4$ in Fig.~\ref{fig.chi}(b).

Since it is difficult to infer the value or even the sign of $J$ from the
paramagnetic Curie temperature, we propose another route to extract $J$. For
$J=h=0$ we have $\chi={2\gamma}/{T}$, cf.\ Eq.~(\ref{2.chiex2}). This
identity does not hold for nonzero $J$ so that the \emph{deviation} from it
might be a good measure of $J$. We thus consider the ratio $\mu \equiv {\chi
T}/{2\gamma}$ at vanishing magnetic field. From Eq.~(\ref{3.z2}) we obtain
\be
\gamma = \frac{1+\partial g/\partial \Delta}{2}
  \cong \frac{5}{6}
  + \frac{5(4V-3\Delta)}{54\,T} + {\cal O}(1/T^2)
\ee
and, at $h=0$,
\be
\mu = \frac{-T\,\partial^2g/\partial h^2}{1+\partial g/\partial \Delta}
  \cong 1 + \frac{10J}{3T} 
  + {\cal O}(1/T^2) .
\ee
The leading high-temperature behavior of $\mu$ is entirely
determined by $J$. This implies that the effect of $V$ and $\Delta$ on the
high-temperature susceptibility only comes from the HS fraction.
Numerical results suggest that $\mu>1$ ($\mu<1$) for $J>0$ ($J<0$) at
all temperatures.
$\chi$ can be measured by standard methods, while the HS fraction
can be obtained independently from optical transmission
experiments\cite{KaM98} or x-ray-absorption near-edge structure\cite{YMK98}
(XANES). The first method makes use of the change in electronic structure
between the LS and HS states, while the second relies on the change of
atomic distances. Since $\chi$ and $\gamma$ are measurable, $\mu$ is a
promising quantity for the determination of the exchange coupling.

\subsection{Finite-temperature crossovers}
\label{sec:cross}

The finite-temperature crossovers found in the preceding subsections are now
studied in more detail.
Regardless of the specific ground state, for high temperatures
the system approaches a limit with $\gamma=5/6$, $\langle m\rangle=0$, and
$w_{\mathrm{eq}}=\gamma^2+(1-\gamma)^2=13/18$. If one starts from the LS
ground state, $\gamma$ has to go from zero to $5/6$. Figures
\ref{fig.gammaLS} and \ref{fig.JgammaLS} suggest that $\gamma(T_\gamma)=1/2$
is a good definition for a crossover temperature $T_\gamma$.
Starting from the alternating phase with $\gamma=1/2$, $\gamma$ first drops
and then recrosses $\gamma=1/2$, at least for small $J$, see
Fig.~\ref{fig.gammaLSHS}.
A measure for the crossover from the alternating phase is
$w_{\mathrm{eq}}(T_w)=1/2$ since $w_{\mathrm{eq}}$ vanishes in the alternating
phase and is close to unity in the HS state at high temperatures.

\begin{figure}[t]
\centerline{\includegraphics[width=3.20in]{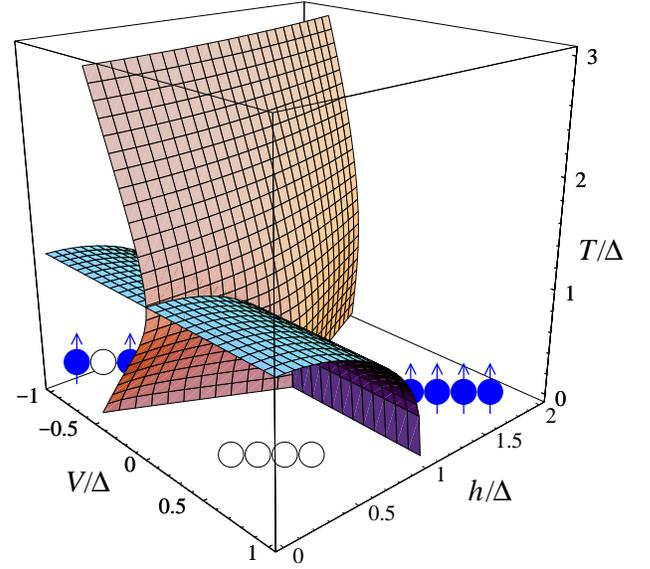}}
\caption{(Color online)
Crossover temperatures for $J=0$. The $T=0$ plane corresponds to
Fig.~\protect\ref{fig.PD0}. The surfaces indicate solutions of
$\gamma=1/2$ and $w_{\mathrm{eq}}=1/2$, respectively.}
\label{fig.PDTw}
\end{figure}

\begin{figure}[tbh!]
\centerline{\includegraphics[width=2.40in]{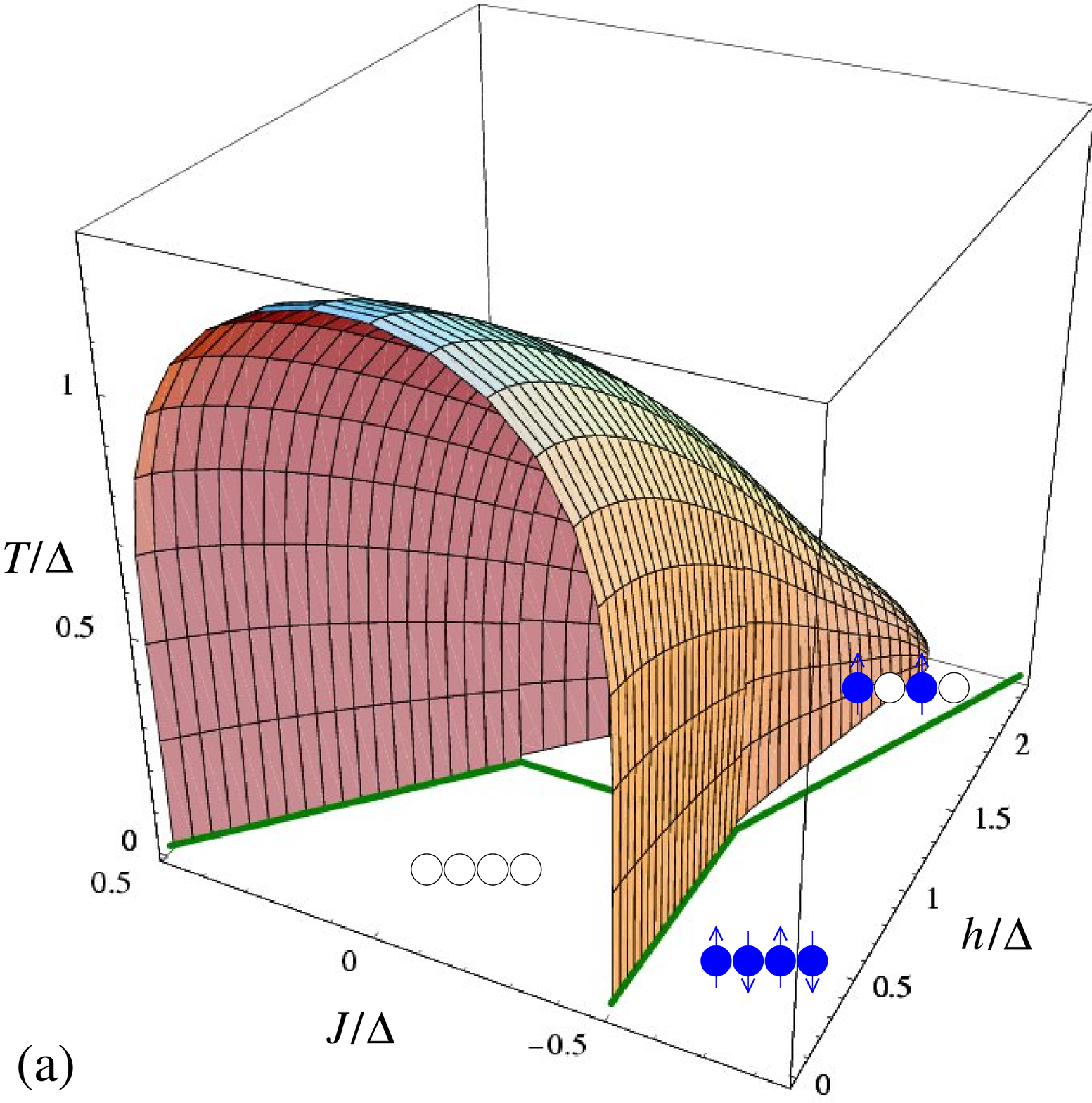}}
\centerline{\includegraphics[width=2.40in]{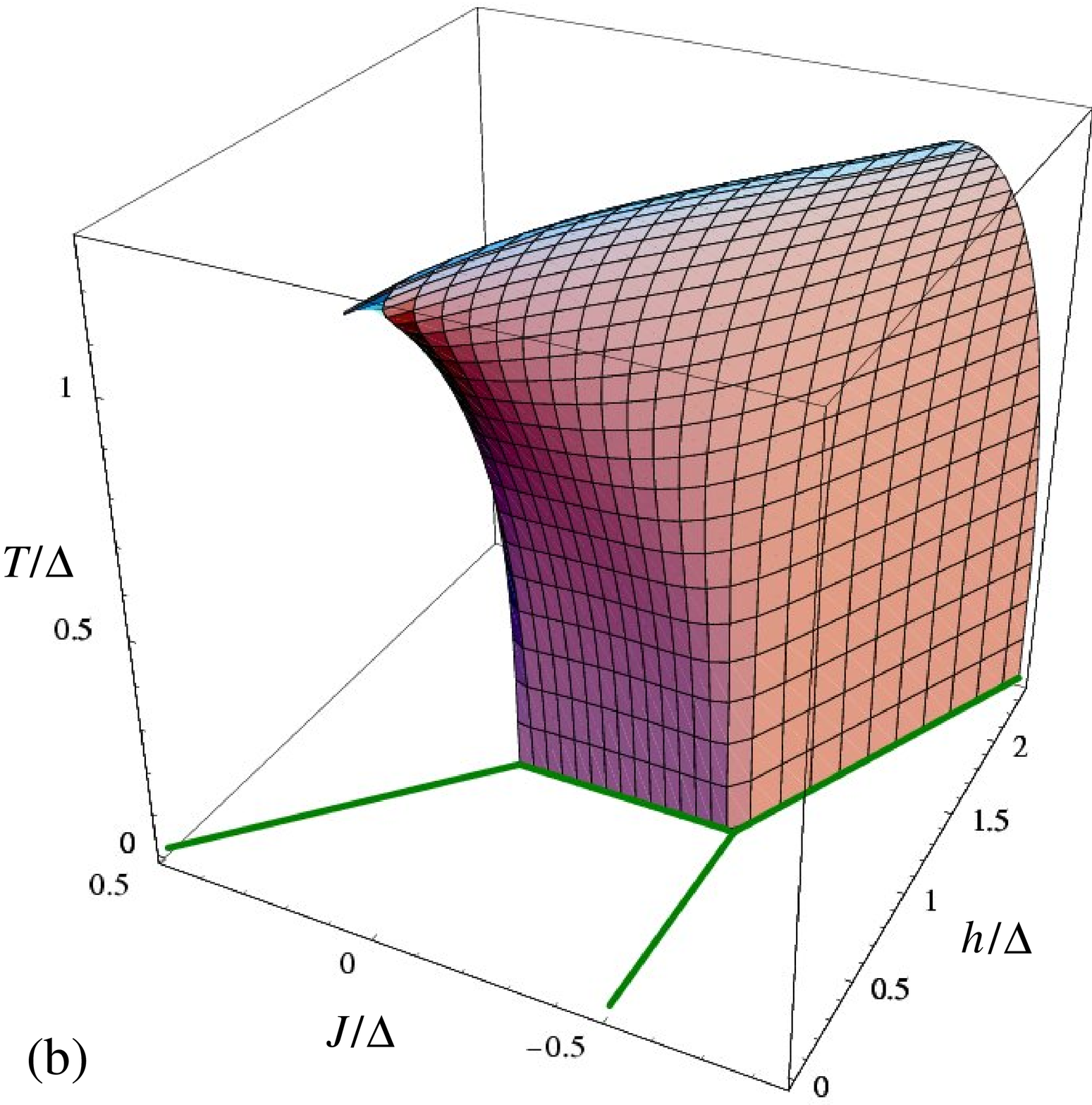}}
\centerline{\includegraphics[width=2.40in]{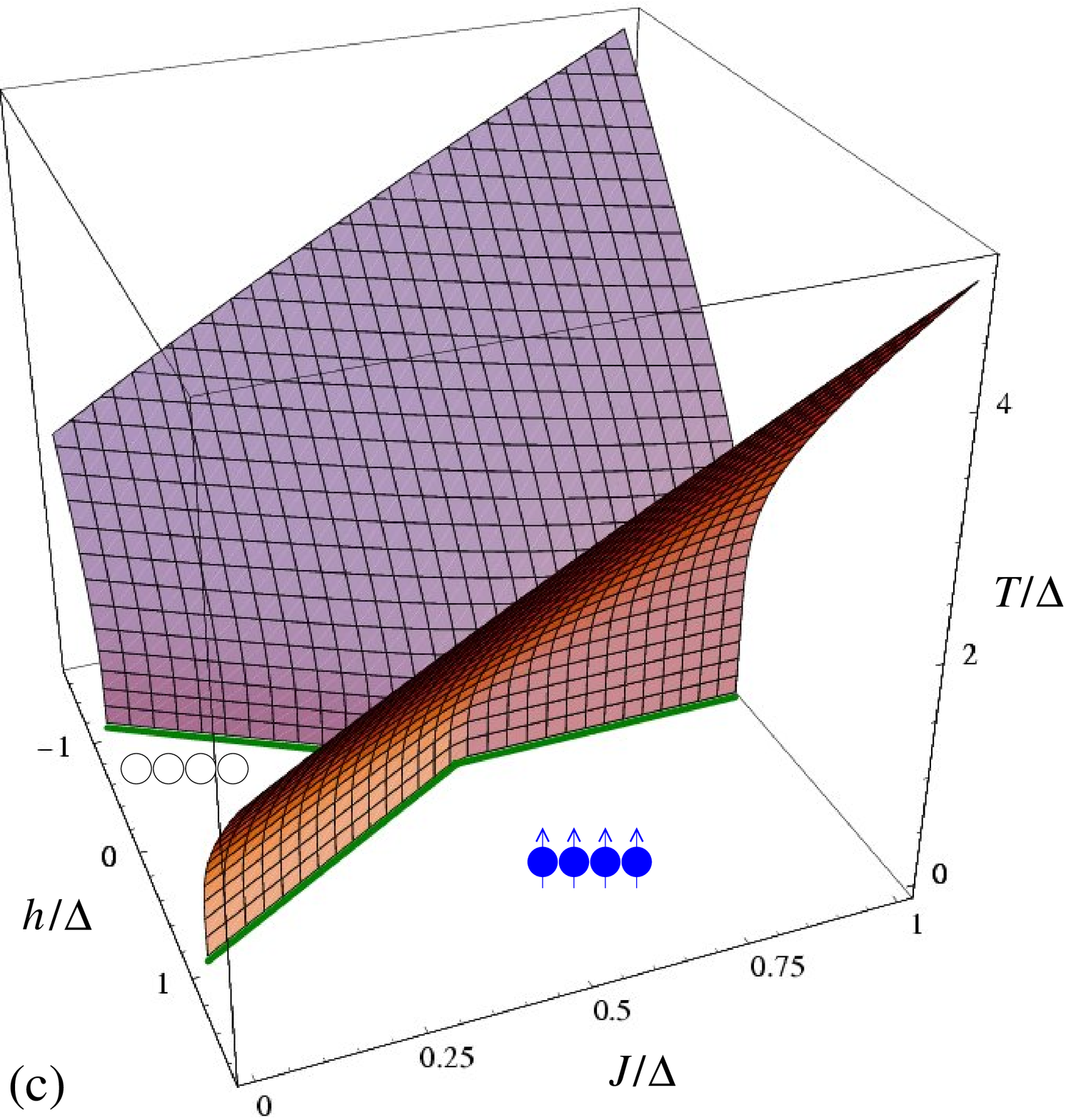}}
\caption{(Color online) Crossover temperatures for $V=0$, defined by
(a) $\gamma=1/2$, (b) $w_{\mathrm{eq}}=1/2$, (c)
$\langle m\rangle/\SHS=1/2$. The $T=0$ plane corresponds to
Fig.~\protect\ref{fig.PD5}. Phase boundaries at $T=0$ are marked by
heavy solid lines.}
\label{fig.PDTJ}
\end{figure}

We first consider the case $J=0$.
Figure \ref{fig.PDTw} shows the crossover temperatures.
From Eq.~(\ref{2.gamma3}),
$\gamma(T_\gamma)=1/2$ is equivalent to $\tilde\Delta=0$. $T_\gamma$ becomes
zero for $|h|/\Delta=2/\SHS=1$. A nontrivial solution exists for
$|h|/\Delta<1$, i.e., for ground states in the LS and part of the
alternating phase. Note that $T_\gamma$ is independent of $V$. The equation
$w_{\mathrm{eq}}(T_w)=1/2$ has \emph{two} solutions (reentrance) if $V<0$
and the ground state is in the LS phase, cf.\ Fig.~\ref{fig.magLS}, whereas
it has only one solution if the ground state is in the alternating phase and
none for $V>0$.
$T_w$ vanishes at the phase boundaries of the alternating ground
state. The LS state
is destroyed by temperatures $T\sim 2\Delta$, the energy per ion for going
from the LS to the HS state. On the other hand, the alternating state is
stable up to $T\sim 2|J|$, the energy cost for breaking LS/HS pairs.

For $J\neq 0$ the typical behavior can already be seen for $V=0$. Figures
\ref{fig.PDTJ}(a,b) show the crossover temperatures defined by $\gamma=1/2$
and $w_{\mathrm{eq}}=1/2$ as functions of $h/\Delta$ and $J/\Delta$. This
should be compared to the phase diagram in Fig.~\ref{fig.PD5}. The reentrant
behavior in $w_{\mathrm{eq}}$ is seen both in Fig.~\ref{fig.PDTw} and
Fig.~\ref{fig.PDTJ}(b).

For the alternating ground state we find that there is (at least) one
nonzero temperature with $\gamma=1/2$ for \emph{some} parameter values, as
in the $J=0$ case. The parameter range where this is the case narrows and
vanishes for fields above $h/\Delta \approx 2.27$, as seen in
Fig.~\ref{fig.PDTJ}(a).
As noted above, we expect significant differences between models with
Ising and Heisenberg exchange if nearest-neighbor pairs in the HS state
appear. This is the case if both $\gamma>1/2$ and $w_{\mathrm{eq}}>1/2$,
i.e., above the crossovers in Figs.~\ref{fig.PDTJ}(a), (b). For the HS
ground states there will be deviations already at small $T$ since the
spin excitations are different in the Ising and Heisenberg models.

The crossover from the ferromagnetic HS ground state to the high-temperature
HS state with $\langle m\rangle=0$ is characterized by the condition
$\langle m\rangle=\SHS/2$. The crossover becomes very sharp at low magnetic
field, cf.\ Fig.~\ref{fig.JgammaHS}. (In the Heisenberg case it might be
broadened by transverse spin fluctuations.) Figure \ref{fig.PDTJ}(c) shows
the crossover temperature for $V=0$, which depends \emph{logarithmically} on
magnetic field for small $h$. This result is known from the 1D Ising model
and can be understood from the formation energy of droplets of reduced spin
polarization.

\section{Summary}

The thermodynamic properties of 1D SCCs with nea\-rest-neigh\-bor elastic
and Ising exchange interactions have been investigated using the exact
transfer-matrix approach. This is motivated by SCCs naturally forming 1D
chains, like $\mathrm{Fe}^{2+}$ with 4-R-1,2,4-triazole ligands, and by
possible artifical 1D structures for molecular electronics. The Ising model
is appropriate for strong easy-axis magnetic anisotropy, for example shape
anistropy in 1D chains. For simplicity, the ratio of the degeneracies of the
LS and HS state of an ion is assumed to be only due to their spin,
corresponding to the assumption of equal vibrational
frequencies.\cite{BoK95,GKA00}

The HS fraction, magnetization, and spin susceptibility are calculated, as
well as the probability of neighboring ions being in the same (LS or HS)
state. These quantities allow one to map out the $T=0$
phase diagram. A strong magnetic field or a ferromagnetic exchange
interaction can drive the system into a ferromagnetic HS ground
state, whereas an antiferromagnetic exchange interaction can lead to an
antiferromagnetic HS ground state. A negative elastic interaction
can drive the system into an alternating LS/HS ground state. These phases
give rise to characteristically different behavior at finite temperatures.
Nonmonotonic temperature dependence of HS fraction and magnetization are
ubiquitous. While there are no phase transitions at $T>0$, the crossovers
between the low-temperature behavior and the HS-dominated high-temperature
limit can become very sharp. Spin crossover chains show 1D analogues of
interesting effects found in the bulk, such as reentrant
transitions\cite{BNM05} and stripe phases.\cite{KhL04}

The susceptibility follows the Curie-Weiss law at high temperatures for all
ground-state phases. The paramagnetic Curie temperature is found to depend
not only on the exchange interaction but also on elastic interaction and
LS/HS energy splitting. It is nonzero even for noninteracting complexes. The
exchange coupling $J$ could be determined experimentally from the
susceptibility $\chi$ and HS fraction $\gamma$ using that the
high-temperature behavior of the ratio $\chi T/2\gamma$ depends only on $J$.

The typical energy scale in all results is $\Delta$, half the LS/HS
energy splitting for a single ion. It would thus be desirable to tune this
scale to small values by a suitable choice of ligands so that
large values of the dimensionless parameters $V/\Delta$, $J/\Delta$,
and $h/\Delta$ can be reached. This would allow to test the rich
temperature-dependent effects predicted here.

\acknowledgments

I would like to thank P. J. Jensen for valuable discussions. Support by the
Deutsche For\-schungs\-ge\-mein\-schaft through Sfb 658 is gratefully
acknowledged.

\end{document}